\documentclass[twocolumn]{aastex63}
\usepackage[utf8]{inputenc}
\usepackage{xcolor}
\usepackage{amsmath}
\usepackage{affils}

\newcommand{\mbh}{\ensuremath{M_\mathrm{BH}}}

\newcommand{\sigmastar}{\ensuremath{\sigma_*}}
\newcommand{\msigma}{\ensuremath{M_{\mathrm{BH}}-\sigmastar}}
\newcommand{\msun}{\ensuremath{M_{\odot}}}
\newcommand{\mstar}{\ensuremath{M_*}}

\begin{document}

\title{A Search for Wandering Black Holes with Gaia and DECaLS}
\author[0000-0002-5612-3427]{Jenny E. Greene}
\affiliation{Department of Astrophysical Sciences, Princeton University, Princeton, NJ 08544, USA}
\author[0000-0002-0041-4356]{Lachlan Lancaster}
\affiliation{Department of Astrophysical Sciences, Princeton University, Princeton, NJ 08544, USA}
\author{Yuan-Sen Ting}
\affiliation{Institute for Advanced Study, 1 Einstein Drive, Princeton, NJ 08540, USA}
\affiliation{Department of Astrophysical Sciences, Princeton University, Princeton, NJ 08544, USA}
\affiliation{Observatories of the Carnegie Institution of Washington, 813 Santa Barbara Street, Pasadena, CA 91101, USA}
\affiliation{Research School of Astronomy \& Astrophysics, Australian National University, Cotter Rd., Weston, ACT 2611, Australia}
\author[0000-0003-2644-135X]{Sergey E. Koposov}
\affiliation{Institute for Astronomy, University of Edinburgh, Royal Observatory, Blackford Hill, Edinburgh EH9 3HJ, UK}
\affiliation{Institute of Astronomy, University of Cambridge, Madingley Rd, Cambridge, CB3 0HA, UK}
\author[0000-0002-1841-2252]{Shany Danieli}
\altaffiliation{NASA Hubble Fellow}
\affiliation{Institute for Advanced Study, 1 Einstein Drive, Princeton, NJ 08540, USA}
\author{Song Huang}
\affiliation{Department of Astrophysical Sciences, Princeton University, Princeton, NJ 08544, USA}
\author{Fangzhou Jiang}
\affiliation{TAPIR, California Institute of Technology, Pasadena, CA 91125, USA}
\author[0000-0003-4970-2874]{Johnny P. Greco}
\altaffiliation{NSF Astronomy \& Astrophysics Postdoctoral Fellow}
\affiliation{Center for Cosmology and AstroParticle Physics (CCAPP), The Ohio State University, Columbus, OH 43210, USA}
\author{Jay Strader}
\affiliation{Center for Data Intensive and Time Domain Astronomy, Department of Physics and Astronomy, Michigan State University, East Lansing, MI 48824, USA}
\date{Feb 2021}

%\maketitle

\begin{abstract}
We present a search for ``hyper-compact'' star clusters in the Milky Way using a combination of Gaia and the Dark Energy Camera Legacy Survey (DECaLS). Such putative clusters, with sizes of $\sim 1$ pc and containing 500-5000 stars, are expected to remain bound to intermediate-mass black holes (\mbh$\approx 10^3-10^5$~\msun) that may be accreted into the Milky Way halo within dwarf satellites. Using the semi-analytic model {\tt SatGen} we find an expected $\sim 100$ wandering intermediate-mass black holes with if every infalling satellite hosts a black hole. We do not find any such clusters. Our upper limits rule out 100\% occupancy, but do not put stringent constraints on the occupation fraction. Of course, we need stronger constraints on the properties of the putative star clusters, including their assumed sizes as well as the fraction of stars that would be compact remnants. 
\end{abstract}

\section{Wandering Black Holes}
\label{sec:intro}

Supermassive black holes are ubiquitous in the centers of massive galaxies today. They may play an important role in regulating star formation in galaxies \citep[e.g.,][]{silkrees1998,mcconnellma2013,kormendyho2013}. We do not know when or how supermassive black holes are formed. Upcoming gravitational wave experiments like \emph{Laser Interferometer Space Antenna} \citep[\emph{LISA}; e.g.,][]{amaro-seoaneetal2015} will potentially be sensitive to the mergers of the first ``seed'' black holes. However, with limited knowledge of both the rates and the mass functions of these black holes, the \emph{LISA} observations cannot uniquely determine either. An important complementary clue will be provided by the study of relic intermediate-mass black holes, with the mass distribution and environment of these black holes today carrying some memory of when and how they were formed \citep[e.g.,][]{volonteri2010,vanwassenhoveetal2010,ricartenatarajan2018,bellovaryetal2019}. The black holes that are found outside of galaxy nuclei may be particularly sensitive to the seeding mechanism \citep[see arguments in][]{greeneetal2020}. 

We expect to find black holes wandering in galaxy halos regardless of seeding mechanism because by hierarchical merging, galaxies are accreted by the Milky Way throughout its history, and some will be totally stripped apart from a stellar nucleus \citep[e.g.,][]{zinneckeretal1988}. Some of these satellites, at least at the massive end, may bring intermediate-mass black holes into the galaxy with them \citep[][]{,volonteriperna2005,bellovaryetal2010}. The central black hole can retain a small cluster of bound stars, which we will refer to as a hyper-compact star cluster \citep[e.g.,][]{merrittetal2009,lenaetal2020}. A related mechanism for depositing off-nuclear black holes is gravitational recoil of black holes from the centers of merger remnants at early times \citep[e.g.,][]{volonterihaardtmadau2003,olearyloeb2009}. Finally, if black holes are made through gravitational runaway at star cluster centers \citep[e.g.,][]{millerhamilton2002,portegieszwartetal2002}, then there may be a population of $\sim 1000$~\msun\ black holes whose surrounding cluster has mostly dissolved, with only a hyper-compact cluster remaining \citep[e.g.,][]{fragioneetal2018}. A primary goal of this paper is to search for these star clusters that would, in turn, point to a population of black hole wanderers. 

We do see evidence that star clusters come into our galaxy with their satellite.  More specifically, nuclear star clusters are massive stellar clusters that are found at the centers of galaxies. It is thought that some fraction of the most massive known star clusters are the remnants of a stripping process that leaves nothing more than the nucleus \citep[see, e.g., ][]{pfefferetal2016,neumayeretal2020}. The cluster M54 is the nucleus of the disrupting Sagittarius dwarf \citep[][]{ibataetal1995}. Another likely stripped nucleus is $\omega$~Cen, which shows evidence for multiple episodes of star formation \citep[e.g.,][]{norrisetal1996,pfefferetal2016}. We have not yet found definitive evidence of intermediate-mass black holes in either cluster \citep{ibataetal2009,noyolaetal2010,lutzgendorfetal2012,baumgardt2017,baumgardtetal2019}, nor for a similar cluster (G1), in Andromeda \citep{gebhardtetal2005}. So far there are no definitive detections of black holes in Milky Way globular clusters \citep[see discussion in][]{greeneetal2020}.

There is, however, evidence in galaxies more massive than the Milky Way for a population of wandering black holes. There are numerous dynamical detections of black holes in ``ultra-compact'' dwarfs \citep{sethetal2014,ahnetal2018,voggeletal2019}. A number of the most massive ultra-compact dwarfs now have dynamically detected black holes at their centers, with the black hole accounting for $\sim 10\%$ of the mass of the system in some cases. As outlined above, the most likely explanation for these objects is that they were formed in a more massive galaxy that was then stripped when falling into its current host halo \citep[e.g.,][]{pfefferetal2014}. 

Perhaps the most compelling case for a wandering black hole is the ``hyper-luminous'' X-ray source HLX1 \citep{farrelletal2009}. HLX1 is an accreting black hole, with a likely mass of \mbh$\approx 10^4$~\msun\ \citep[e.g.,][]{davisetal2011,webbetal2012}. It is found in a cluster of stars \citep{farrelletal2014} sitting a few kpc from a more massive galaxy ESO 243-49 at $z = 0.022$, which is very likely the remnant of a stripped dwarf galaxy that was accreted by the more massive system. 

Our goal here is to search for lower-mass stellar clusters that would be bound because of the presence of an intermediate-mass black hole. \citet{olearyloeb2012} and \citet{lenaetal2020} investigate a search  for hyper-compact star clusters using stellar colors. In this work, we instead focus almost exclusively on spatial clustering information. As described in \S \ref{sec:prediction}, we expect any clusters deposited from accreted satellites to be found relatively near to the Galactic Center ($<50$ kpc) and thus to be rather large on the sky and easily resolvable with existing ground-based imaging surveys. Therefore, in \S \ref{sec:data} and \S \ref{sec:search}, we describe a joint search using the spatial resolution of Gaia and the depth of the Dark Energy Camera Legacy Survey \citep[DECaLS;][]{deyetal2019}. In \S \ref{sec:limits}, we summarize the limits we derive from our non-detections and in \S \ref{sec:summary}, we consider the future prospects of this work. 

\section{Expected Properties of the Wanderers}
\label{sec:prediction}
Our goal is to search for the hyper-compact stellar clusters that should accompany an intermediate-mass black hole wandering through our galaxy. In this section we summarize the relevant theoretical understanding of the size, mass, and stellar content of these hypothetical objects, so that we can hone our search strategy. It is worth emphasizing that we tune our search parameters to the specific case of ``ex-situ'' wanderers that formed in an external dwarf galaxy and were subsequently accreted by the Milky Way. In order for the galaxy to be fully stripped, we expect such clusters to live relatively close ($\lesssim 50$~kpc) to the Galactic Center based on modeling presented in \S \ref{sec:satgen}.

Numerous successful searches in modern wide-field imaging surveys for globular clusters, dwarf galaxies, and stellar streams have been carried out in the Milky Way over the past two decades \citep[e.g.,][]{willmanetal2005,belokurovetal2006,koposovetal2008,bechtol2015,koposovetal2015,drlica-wagner2015,shipp2018,torrealbaetal2019}.  However, no search has been tuned to the compact sizes and low numbers of stars that we believe the hyper-compact star clusters may contain. Therefore, we thought it worthwhile to perform a customized search matched to the small scales of our target population.

\subsection{Predicted size and mass}
\label{sec:olearyloeb}

The most detailed calculations of the dynamical evolution of bound remnant clusters have been made by \citet{olearyloeb2009}, followed by confirming simulations in \citet{olearyloeb2012}. These authors focus on wanderers formed through very early mergers, in which the merged remnant is ejected from the proto-Milky Way center via gravitational wave recoil \citep[e.g.,][]{peres1962,campanellietal2007}, a model also explored by \citet{volonteriperna2005}. Their modeling of the subsequent dynamical evolution of the cluster is very likely relevant to the final mass and size of the star clusters we consider as well, although the mechanism for forming the wanderers is different. 

If we assume that the black hole is able to retain roughly its mass in stars within its sphere of influence, then the calculations of \citet{olearyloeb2012} suggest that by the present day, the cluster will lose roughly 60-80\% of that mass through relaxation \citep[see arguments in][]{rashkovmadau2014}. At the same time, the clusters will grow in size as $t^{1/3}$, leading to present-day clusters of 0.5-1 pc in size, with mass $\sim 20\%$ of the black hole mass.

We highlight two major potential caveats here. First, there is the possibility that the compact cluster is dominated not by visible stars, but rather by stellar-mass black holes. Such compact remnants preferentially reside at the cluster core and they would not be visible by electromagnetic means \citep[e.g.,][]{baumgardtetal2019,gielesetal2021}. \citet{olearyloeb2012} acknowledge this possibility as well, but do not pursue it. We will take the same approach, but also note that extreme mass-ratio inspiral events, the detection of the merger of a stellar-mass black hole with an intermediate-mass black hole, would be one way to detect such clusters in the future with \emph{LISA} \citep[e.g.,][]{gairetal2010}.

The second caveat is the possibility that there is no stellar cusp around the black hole. \citet{bahcallwolf1976} calculate the stellar distribution around a black hole that ensues when the black hole is embedded in a stellar cluster with much higher mass than the black hole. However, low-mass galaxies can have very low stellar densities if they are not nucleated, and while the nucleation fraction of $\sim 10^9$~\msun\ galaxies is near unity, that fraction drops substantially at lower galaxy stellar mass \citep[e.g.,][]{sanchez-janssenetal2019,neumayeretal2020}. Therefore, it is possible that in practice, the wandering black holes lack even the hyper-compact clusters that we consider here. To partially compensate for this possibility, we consider the case that only nucleated galaxies host observable wandering black holes in \S \ref{sec:satgen}.

%%%%%%%%%%%%%%%%%%%%%%%%%%%%%%%%%%%%%%%%%%
\begin{figure*} %[ht!]% 
\hspace{-5mm}
\includegraphics[width=9cm]{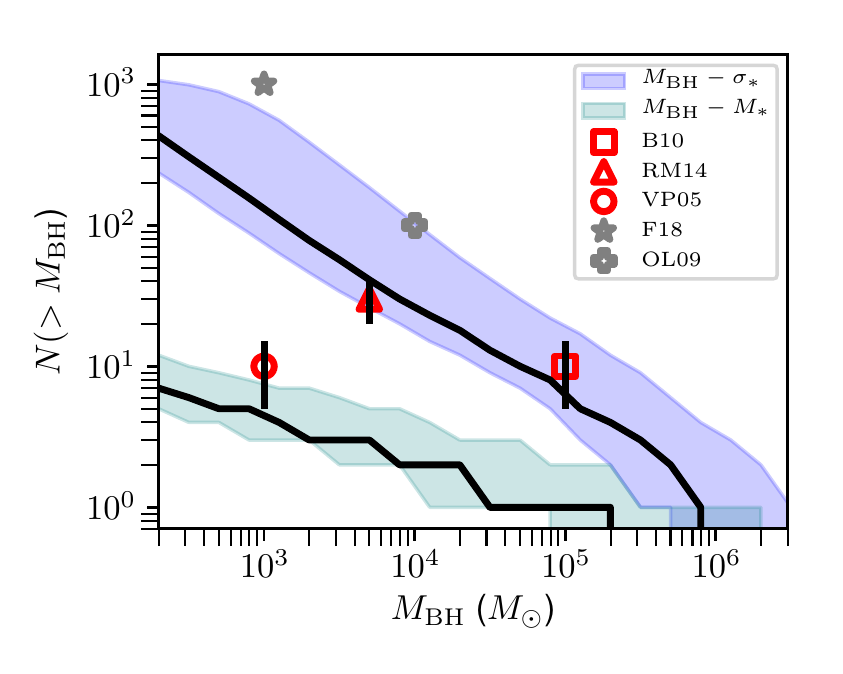}
\hspace{-3mm}
\includegraphics[width=9cm]{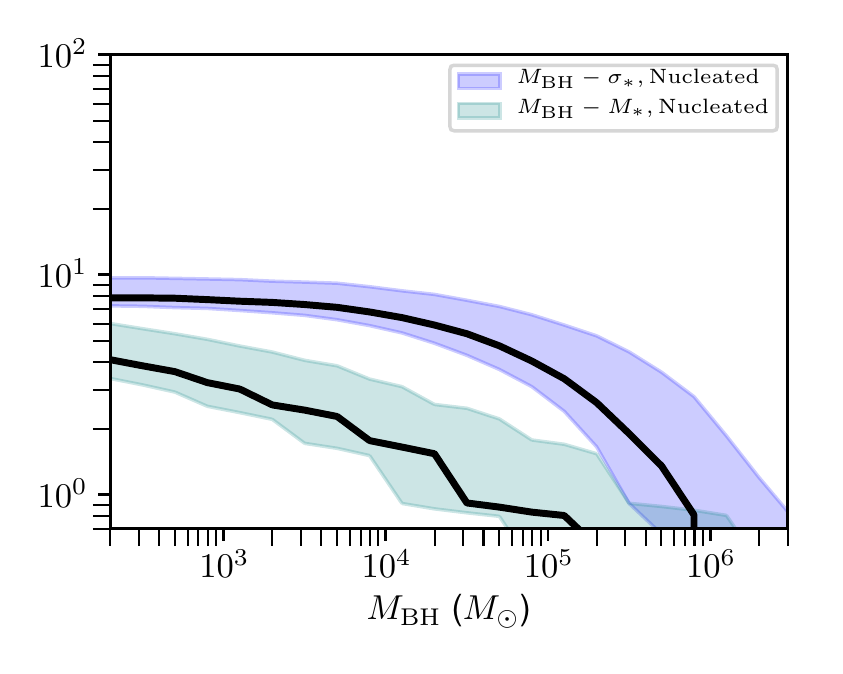}
\vspace{-2mm}
\caption{{\it Left}: The cumulative number of black holes predicted assuming that the only channel to make wanderers is through accreted satellites. We estimate the central \mbh\ of each satellite based either on the stellar mass (teal) or velocity dispersion (blue) at the time of accretion; the scatter includes both the range of accreted satellite masses in the {\tt SatGen} models and scatter in the scaling relations. We show the $1 \sigma$ range. As these are both extrapolations, they predict very different mass functions for the same population of disrupted satellites. We assume 100\% occupation fraction in this panel, which is thus an upper limit on the number of wanderers formed via accretion of satellites. Roughly speaking, the clusters should contain $\sim 20\%$ of the mass of the black hole \citep{olearyloeb2012}. 
For comparison, we plot predictions from prior work considering satellites \citep[red symbols; VP05, B10, RM14][]{volonteriperna2005,bellovaryetal2010,rashkovmadau2014} as well as predictions from formation in globular clusters that are then disrupted \citep[grey star;][]{fragioneetal2018} or through recoil \citep[grey plus][]{olearyloeb2009}. These are cumulative, so we plot them at the lowest black hole mass considered. 
{\it Right}: Same calculation as at left, but in this case we take the nucleation fractions from \citet{sanchez-janssenetal2019} and assume that only nucleated galaxies contain black holes. }
\label{fig:nbh}
\vspace{1pt}
\end{figure*}
%%%%%%%%%%%%%%%%%%%%%%%%%%%%%%%%%%%%%%%%%%

%%%%%%%%%%%%%%%%%%%%%%%%%%%%%%%%%%%%%%%%%%
\begin{figure} %[ht!]% 
\hspace{-5mm}
\includegraphics[width=9.5cm]{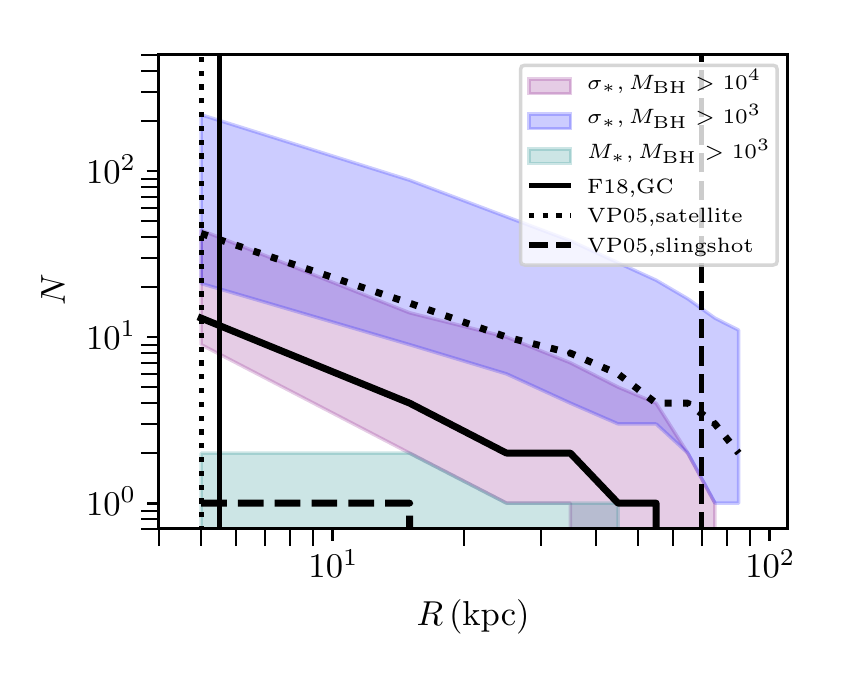}
\vspace{-5mm}
\caption{Radial distributions of clusters that host black holes with \mbh$>10^3, 10^4$~\msun\ where black hole mass is estimated following the \msigma\ relation (blue-dotted and purple-solid respectively) or the $M_{\rm BH}-M_*$ relation (teal long-dash, all black holes with \mbh$>10^3$~\msun). We also schematically indicate the typical radius of black holes made in recoil events from early mergers or from accreted satellites from \citet{volonteriperna2005}. We highlight that wanderers deposited by disrupted globular clusters would be very close to the Galactic Center \citep{fragioneetal2018}, and we note that \citet{olearyloeb2012} do not present a radial distribution, but do predict that black holes will extend to 100 kpc. }
\label{fig:radbh}
%\vspace{1pt}
\end{figure}
%%%%%%%%%%%%%%%%%%%%%%%%%%%%%%%%%%%%%%%%%%

\subsection{Predicted number density and radial distribution}
\label{sec:satgen}

There have been a number of estimates for the number of wandering black holes that we might expect to find in a Milky Way-like galaxy \citep[e.g.,][]{volonterihaardtmadau2003}. We update these predictions in two ways. First, we utilize updated scaling relations that account for both the available data and upper limits \citep{greeneetal2020}. There is considerable scatter in these relations, which are of course extrapolations to the mass regime of interest here. Nevertheless, new dynamical measurements and constraining upper limits from the centers of nearby low-mass galaxies \citep{neumayerwalcher2012,nguyenetal2018,nguyenetal2019b} motivate us to recalculate estimates of wanderer number density.

Second, considerable work on high-resolution hydrodynamical simulations have yielded a new generation of semi-analytic models that can simultaneously capture the evolution of satellites in the tidal field of their host, and allow us to examine a large suite of Milky Way-like models in contrast with individual high-resolution simulations that have been used for this purpose in the past \citep[e.g.,][]{rashkovmadau2014}. Specifically, we use the semi-analytic model {\tt SatGen} \citep{jiangetal2020}. 
This model builds upon Monte-Carlo dark matter halo merger trees and follows the evolution of satellites using tidal-evolution tracks calibrated against high-resolution idealized N-body simulations.
The model also takes into account the response of dark matter halos to baryonic feedback, as formulated from zoom-in cosmological hydro-simulations, and the gravitational influence of a galactic disk on satellite evolution. In this way, it is possible to generate a large suite of Milky-Way-mass systems emulating those from high-resolution zoom-in cosmological simulations regarding satellite statistics, with a more complete sampling of the halo-to-halo variance and at numerical resolutions comparable to or higher than that of the simulations. 

We use {\tt SatGen} models with present-day halo masses distributed uniformly in the range $M_{\rm halo} = 10^{12}-10^{12.3}$~\msun. These models include an evolving disk component and are made to emulate simulations of bursty stellar feedback such as the FIRE and NIHAO simulations. Satellites with halo masses before infall that are $M_h> 1.3 \times 10^6$~\msun\ are considered.

From the model, we are particularly interested in satellites that dissolve, which is assumed to happen in {\tt SatGen} when the sub-halo is stripped to a mass of $10^6$~\msun. As shown by Jiang et al., a large fraction of the most massive satellites, particularly those accreted early, do pass close enough to the host halo center to be dissolved. Of course, it would be very interesting to follow these systems dynamically, including the impact of a black hole and possible stellar cluster on the subsequent stellar content and dynamical friction, but that is beyond the scope of the present work. For now, we simply count the number, infall mass, and radial distribution of these disrupted satellites (Figure \ref{fig:nbh} \& Figure \ref{fig:radbh}). 

Dissolved halos are still tracked as point masses by {\tt SatGen}. We exclude halos that are within 10 pc of the galaxy center, assuming these effectively have merged with the central black hole. In practice such occurrences are very rare, although the satellite orbits are drawn from a cosmological distribution neglecting mass, so there is some chance that {\tt SatGen} underestimates cannibalism if more massive satellites are preferentially on more radial orbits.

To convert the observed halo and stellar properties at infall into estimated black hole masses, we must decide what fraction of satellite halos will host black holes. The so-called ``occupation fraction'' of black holes in dwarf galaxies is not yet well measured. However, given the upper limit on a black hole in the nearby dwarf galaxy M33 \citep{gebhardtetal2001}, it is likely that not every dwarf galaxy hosts a massive central black hole. We first calculate the predicted numbers of black holes under an assumption of 100\% occupation, but additionally calculate the expected number of wanderers if we were to assume that only nucleated galaxies (galaxies with nuclear star clusters) host black holes. We adopt the nucleation fraction from \citet{sanchez-janssenetal2019}, specifically their measurements in the Virgo Cluster. Virtually all galaxies with $M_* \approx 10^9$~\msun\ host nuclei, falling with mass such that galaxies with $M_* < 10^6$~\msun\ will host black holes $<10\%$ of the time.

We compute two possible black hole masses. First, we use the stellar mass at infall from the model combined with an \mbh-$M_*$ relation from \citet{greeneetal2020}. There are two major uncertainties here. First, we have assumed a stellar-to-halo mass relation to assign each halo a stellar mass \citep[in this case the relation from][]{rodriguez-pueblaetal2017}. The relationship between stellar and halo mass is notoriously unconstrained at these dwarf masses, with large degeneracy between the scatter and slope of the relation \citep[e.g.,][]{munshietal2021}, leading to large uncertainty in the stellar masses within the {\tt SatGen} model. Second, the relationship between stellar mass and black hole mass has considerable intrinsic scatter and is dependent on galaxy morphology at higher mass \citep[e.g.,][]{reinesvolonteri2015}. We are extrapolating a high-scatter relation into an unknown regime, which adds considerable systematic uncertainty to these estimates. 

As a second \mbh\ estimate, we take the circular velocity at infall and, following \citet{rashkovmadau2014}, we calculate $\sigmastar = v_{\rm circ}/2.2$ and then use the \mbh-$\sigmastar$ relation \citep{greeneetal2020}. In this case, the circular velocity is securely predicted from the model, but the conversion between circular velocity and stellar velocity dispersion is not well known in the dwarf regime. As far as the scaling relations, the \msigma\ relation is also somewhat morphology dependent \citep[e.g.,][]{greeneetal2016}, but the scatter is lower than the conversion based on total stellar mass. We will take the $v_{\rm max}$-based predictions presented below as the primary predictions throughout the paper, because at least we will not be directly dependent on an assumed stellar-to-halo mass relation. At the end of this section, we will briefly discuss a third model for black hole mass in which there is no scaling with galaxy properties and all black holes are relatively massive at birth.

In calculating the expected number of wandering black holes, we include the halo-to-halo scatter by using the 85 Jiang et al.\ Milky Way-like models. We also vary the mapping between halo and black hole mass, by drawing 100 black holes per satellite from within the published scatter in the scaling relations. The resulting cumulative numbers of predicted wanderers are shown in Fig.\ \ref{fig:nbh} for the \sigmastar\ and \mstar\ scalings respectively. 

Roughly speaking, the stellar mass in the clusters will be $\sim 20\%$ of \mbh. Thus, we limit our attention to black holes $>10^3$~\msun, where we might hope to detect the bound stellar cluster. Under an assumption of full occupation, we expect to find a few to 100 clusters with a hundreds to a few thousand stars, depending on both the black hole scaling relation adopted and on the occupation fraction of black holes in the infalling satellite population. 

We have an estimate of the satellite stellar masses from a stellar-to-halo mass relation \citep{rodriguez-pueblaetal2017}. Assuming black holes with \mbh$>10^3$~\msun, in the case of the \mbh-$M_*$ relation, nearly all hosts will have $M_* > 10^7$~\msun. In contrast, in the \sigmastar-based scaling, much lower-mass satellites host black holes. Specifically, we find that galaxies with stellar mass $>10^5$~\msun\ are predicted to host black holes when scaling with \sigmastar. Thus, we find many more possible black holes in the \sigmastar-based scaling. Also, we note that even if we were to extend the lower-limit on halo masses considered by the model, we would not find any more black holes with \mbh$>10^3$~\msun.

Depending on the scaling relation, we not only expect different stellar masses, but also a different radial distribution, of satellites hosting black holes, as shown in Fig.\ \ref{fig:radbh}. The most massive satellites, those that preferentially host black holes in the \mbh-$M_*$ case, must be accreted early and travel close to the galaxy center to be stripped \citep{jiangetal2019}. In contrast, the much wider range of halo and stellar mass that can host black holes when scaling with \sigmastar\ also translates to a wider radial range, $R \lesssim 50$~kpc.  We will use the larger distance limit in \S \ref{sec:limits} when we calculate upper limits on the number of wanderers in the Milky Way.

In Figure \ref{fig:nbh}, we compare with existing similar predictions for the number of wandering black holes in a Milky Way-mass halo. Each paper makes slightly different assumptions about seeding mechanisms, and thus predict different mass spectra for the resulting wanderer population. \citet{volonteriperna2005} use a semi-analytic model as well, and directly consider multiple seeding mechanisms, to predict between one and ten black holes with \mbh$>10^3$~\msun. \citet{bellovaryetal2010} predict slightly higher numbers (5-15), in this case with a heavy seeding model. \citet{rashkovmadau2014}, like us, have no seeding model, but instead assign \mbh\ based on properties of the halos, and predict comparable cumulative numbers as we do, using a similar \mbh-\sigmastar\ relation based on peak maximum halo velocity. Both \citet{tremmeletal2018} and \citet{ricarteetal2021} find $\sim 5-10$ wandering black holes in Milky Way-mass halos using the Romulus simulation. Our predictions, and those from the literature, span two orders of magnitude in number because we are extrapolating assumptions about black hole scaling relations and occupation fractions into an unknown regime. These are the primary systematic uncertainties in our predictions, rather than detailed assumptions in the cosmological models that we use.

The predicted radial distributions in prior work are also similar to ours \citep[such as][]{bellovaryetal2010,rashkovmadau2014}, particularly those that model wandering black holes from disrupted satellites alone. The distribution is more extended than the wandering black holes expected from the dissolution of globular clusters should black holes form efficiently in their centers \citep[e.g.,][]{fragioneetal2018}, but is more centrally concentrated than models like \citet{volonteriperna2005} or \citet{olearyloeb2012} that also include a component from early recoil events. In this work we do not attempt to search for more distant (spatially unresolved) clusters, which would require a different approach.

Finally, it is interesting to consider how our predictions would change if we adopted a heavy seeding prescription with a lower-mass limit of $10^4-10^5$~\msun, as might be expected in some heavy seeding models \citep[e.g.,][]{bellovaryetal2010,inayoshietal2020}. If we assume that every satellite, regardless of mass, is seeded with a heavy seed, then the number of expected black holes with \mbh$>10^3$~\msun\ would be very high ($\sim 1000$) and would become very sensitive to our halo mass limit. Such a model is very easily ruled out, as we will show in the following sections. If, on the other hand, we assume a reasonable drop in occupation fraction with mass, then the number and mass distribution of black holes under a heavy seeding model would be very similar to Figure \ref{fig:nbh} ({\it right}), since the occupation fraction becomes a strong function of stellar mass, and this factor matters more than exactly how we assign black hole masses.

%%%%%%%%%%%%%%%%%%%%%%%%%%%%%%%%%%%%%%%%%%
\subsection{Predicted stellar content}
\label{sec:artpop}

We use the \texttt{ArtPop} software package (\citealt{Danieli2018}; \citealt{grecoetal2021}; J. Greco \& S. Danieli, in preparation) to generate model clusters for calibration and completeness analyses. Using the MIST isochrones, we synthesize stellar populations of a given age and metallicity, and simulate realistic images of stellar systems based on their physical and structural parameters. In this work, we use ArtPop models to generate color-magnitude-position predictions in Gaia and DECaLS for expected clusters.

To build our model clusters, we assume a standard \citet{kroupa2001} initial mass function, with slope $\psi \propto m^{-\alpha}$ with $\alpha = 1.3$ for stars below $0.5$~\msun, and $\alpha = 2.3$ at higher stellar mass. We set the lower and upper mass limits based on the MIST isochrones of a given age and metallicity. As described in \S \ref{sec:olearyloeb}, we do not explore the possibility that the present-day mass function is dominated by stellar-mass black holes. We assume a \citet{plummer1911} profile with a scale radius of 1 pc; the clusters are expected to grow to this large size with time \citep{olearyloeb2012}. We assume a fixed size for our mock clusters. If in reality they are considerably larger (or smaller) by more than a factor of two, then our search is unlikely to find them. 

We assume the clusters are 10 Gyr old, similar to their likely accretion time. To span the possible metallicity range of such clusters, we take [Fe/H]$=-1.5, -0.5$, which is measured for nuclear star clusters in host galaxies with $M_* = 10^8-10^9$~\msun\ \citep{neumayeretal2020}. In practice, the nuclei of such low-mass galaxies accreted at $z>1$ are likely to have even lower metallicities, but we conservatively adopt these limits since the hotter and brighter stars in clusters at lower metallicity are easier to find at fixed stellar mass. 

\begin{figure*}[t!]
    \centering
    \includegraphics[width=0.75\textwidth]{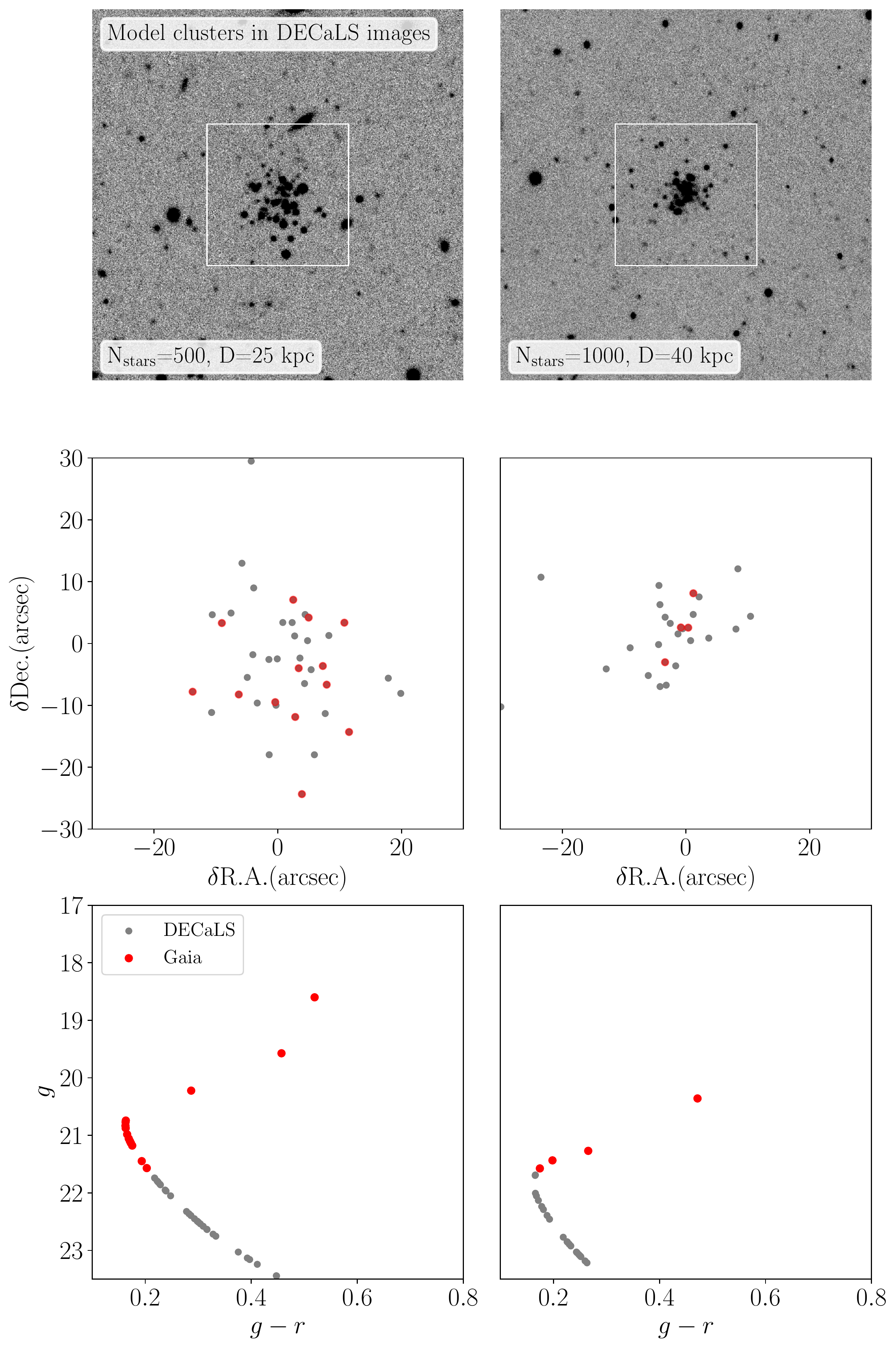}
    \caption{
    {\it Top:} Two model hyper-compact clusters injected into DECaLS images with $N=500$, $\rm{D}=25$~kpc (left) and $N=1000$, $\rm{D}=40$~kpc stars (right). These clusters should be comfortably detected by our Gaia search (see \S \ref{sec:data}).
{\it Middle:} The spatial distribution of the stars in these clusters as seen by Gaia (red) and DECaLS (grey).
{\it Bottom:} Noiseless color-magnitude diagram for the clusters, colors as above. We should note that the number of giant stars for these low-mass clusters will vary quite significantly, from near zero up to fifteen for the $N=1000$ star cluster, with a median of five. 
    }
    \label{fig:modelclust}
\end{figure*}

In Figure \ref{fig:modelclust}, we show the color-magnitude diagrams (bottom panel) and spatial distributions (middle panel) of example model clusters at the limits of our detection threshold (\S \ref{sec:completeness}), with stars detectable in Gaia and DECaLS in red and grey, respectively. The upper panel shows the artificial clusters injected into DECaLS images. It is worth noting that there is a large amount of stochasticity in the number of giant stars for a $N=1000$ star cluster. While the median number of expected stars is five, it can range from zero to 15. We are not complete beyond the distance where we lose main-sequence stars (\S \ref{sec:completeness}) for these low-$N$ models, and so we are not very sensitive to this stochasticity here. But, future searches at larger distances would be unless they are deep enough to reach below the main-sequence turn-off.

%%%%%%%%%%%%%%%%%%%%%%%%%%%%%%%%%%%%%%%%%%%%%%%%%%
%%%%%%%%%%%%%%%%%%%%%%%%%%%%%%%%%%%%%%%%%%%%%%%%%%

\section{Data}
\label{sec:data}

Having defined the parameters of the expected hyper-compact clusters surrounding the putative intermediate-mass black holes, we now turn to search for them. We first describe the data sets, and then in \S \ref{sec:search} we describe our search.

\subsection{The Gaia catalog}
\label{subsec:gaia_count_catalog}

Gaia is a European Space Agency astrometry mission. Here we use the Early Data Release Three of the Gaia mission \citep[Gaia EDR3;][]{gaia2016,gaia2018,gaia2021} to make an initial search for stellar overdensities on $\sim 15$~\arcsec\ scales by looking for stars with anomalously large numbers of neighbors on this scale. To do this, we utilize a custom-built catalog from S.~Koposov\footnote{The catalog has been built based on the vanilla EDR3 Gaia source catalog using the Whole Sky SQL Database maintained in Cambridge using the Q3C spatial query software \cite{koposovbartunov2006}.} based on the EDR3 that records the number of neighbors that each source has with angular separation less than $x$ arcsec for 10 aperture choices of $x$. The apertures range from 0.25 to 128 arcsec, with each aperture increasing by a factor of two. It is convenient to work with the counts within ``annuli" (i.e. number of neighbors between $y$ and $x$ arcsec), since each annulus can be modeled independently. We will refer to the number of neighbors between $x/2$ and $x$ arcec as annuli counts $A_x$.  Our fundamental data set is therefore a vector of counts 
\begin{equation}
    \label{eq:Ndef}
    \mathbf{A} \equiv \left(A_{0.25}, A_{0.5}, ...
    A_{64}, A_{128}\right)
\end{equation}
within these apertures for each EDR3 source.

\subsubsection{Masking of known galaxies and star clusters}

There are two main types of contaminant that we mask before performing our cluster search. First, we mask all known groupings of stars in the Milky Way, including open clusters, globular clusters, and known satellites compiled by \citet{Torrealba18} from several different sources \citep{McConnachie12,Harris10}. In general we mask a two degree radius around each satellite, except in the cases of the Large and Small Magellanic Clouds, along with Fornax, which we mask with a 10 degree radius (while the Large and Small Magellanic Clouds themselves do not fall within the DECaLS footprint, their outskirts turn out to be a major contaminant). 

We also find that the cores of nearby galaxies can appear in Gaia as a set of point sources, and so we also mask the New General Catalog (NGC) galaxies\footnote{https:\///github.com\//mattiaverga\//OpenNGC\//blob\//master\//NGC.csv}. The majority of these galaxies are given a 1-2\arcmin\ radius mask, while a small subset of galaxies with larger tabulated radii\textbf{} are masked over a full degree radius. In all, we mask roughly 1000 deg$^2$ of the 9000 deg$^2$ area covered by DECaLS.

\subsection{DECaLS}

The DECam Legacy Survey (DECaLS) has been carried out with the Dark Energy Camera \citep[DECam;][]{flaugheretal2015,deyetal2019} at the Mayall 4m telescope at the Cerro Tololo Inter-American Observatory. DECaLS reaches $5 \sigma$ point-source depths of $grz = 23.95, \,23.54,\, 22.50$ AB mag over 9000 deg$^2$. We utilize $gr$ photometry from Data Release 9 of DECaLS\footnote{https:\///www.legacysurvey.org/dr9} in our search.

\section{Search with Gaia+DECaLS}
\label{sec:search}

As described in \S \ref{sec:prediction}, we seek hyper-compact star clusters comprising 500-5000 stars within $\sim 1$ pc. These clusters likely fall within 50 kpc of the Galactic Center (see Fig. \ref{fig:radbh}). In the proof of concept search that we present here, we have decided to combine the spatial resolution of \emph{Gaia} with the depth of DECaLS to search efficiently for possible clusters. First, we use the Gaia EDR3 catalog to search for clusters of $N>4$ stars within $\lesssim 15 \arcsec$, as expected for these hyper-compact star clusters. The benefits of Gaia include the high spatial resolution and the potential to filter on proper motions. However, with a depth of $G \sim 21.5$ mag \citep[][]{gaia2021}, we will detect only a few of the stars with Gaia (as we see in Fig. \ref{fig:modelclust}).

\begin{figure*} %[ht!]% 
\hspace{-5mm}
\includegraphics[width=9cm]{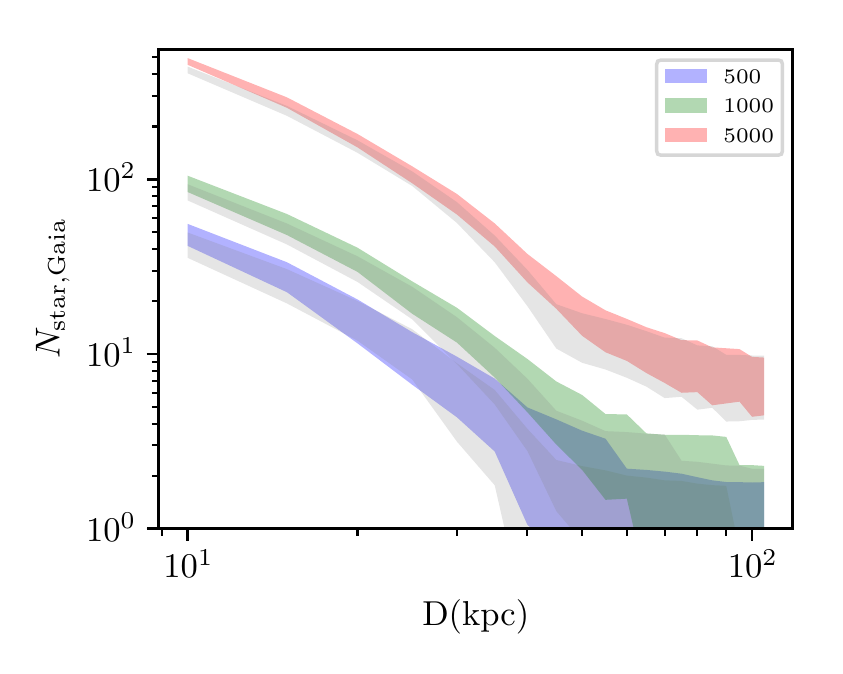}
\hspace{+0cm}
\includegraphics[width=9cm]{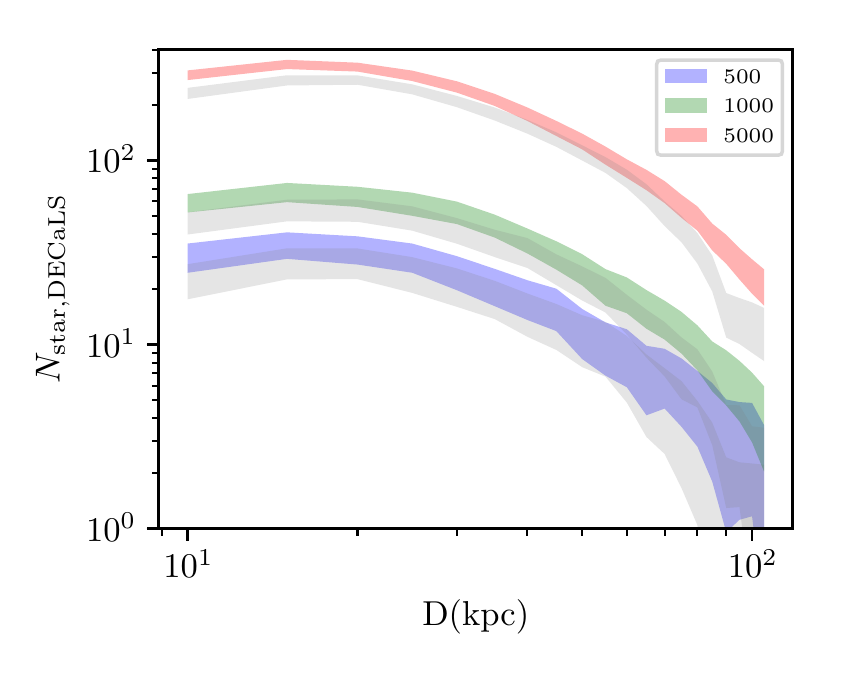}
\vspace{-2mm}
\caption{{\it Left:} The number of detectable stars in clusters with $N=500,1000,5000$ total stars (in blue, green, red respectively) as a function of distance from the Sun. The grey shaded regions are the [Fe/H]$=-0.5$ models, showing the small difference introduced over this modest metallicity range. We will show (\S \ref{sec:completeness}) that when there are more than $\sim 4$ stars, we can reliably identify the cluster in the Gaia data. {\it Right:} Same as left, but for DECaLS imaging with a conservative magnitude limit of $g=23.5$ mag and only considering stars within 15\arcsec\ of the cluster center. }
\label{fig:nstar}
\vspace{1pt}
\end{figure*}

To illustrate this point, in Figure \ref{fig:nstar}, we show the expected number of stars that can be observed with Gaia and DECaLS in a cluster of 500-5000 stars as a function of distance. We use 100 different realizations of the cluster to include stochasticity in how many stars populate the red giant branch for the lower-mass clusters. For $N = 1000$, we expect a large range in the number of giant stars, between zero and twelve, with a median of five. However, over the distance range where we are sensitive with this Gaia search (\S \ref{sec:completeness}), we do reach below the main-sequence turn-off, meaning that our overall $N_{\rm star, Gaia}$ is dominated by main-sequence stars.

The grey regions show the numbers for the higher-metallicity clusters, showing that the range in metallicity leads to only a small difference in the number of stars, so throughout we will focus on models with [Fe/H]$=-1.5$, which seems more likely. We see that overdensities of stars identified with Gaia should be accompanied by an increase of three to five times in the number of stars at detected at DECaLS depth, if we are identifying star clusters. This will be true even with some mild crowding, as discussed in \S \ref{sec:crowding}. 

In this section, we first present the model that we use to identify possible clusters as outliers in spatial clustering (\S \ref{sec:negbin}). We then calibrate the methods using artificial clusters (\S \ref{sec:outlier}), argue that crowding is unlikely to greatly impact our search (\S \ref{sec:crowding}), and determine our completeness (\S \ref{sec:completeness}). Finally, with a list of cluster candidates in hand from Gaia, we use DECaLS to invalidate most candidates based their total number of stars at DECaLS depths (\S \ref{sec:decalsfilter}) along with their colors (\S \ref{sec:cmdfilter}).

%%%%%%%%%%%%%%%%%%%%%%%%%%%%%%%%%%%%%%%%%%
%%%%%%%%%%%%%%%%%%%%%%%%%%%%%%%%%%%%%%%%%%

%%%%%%%%%%%%%%%%%%%%%%%%%%%%%%%%%%%%%%%%%%
\begin{figure} %[ht!]% 
\hspace{-4mm}
\includegraphics[width=9cm]{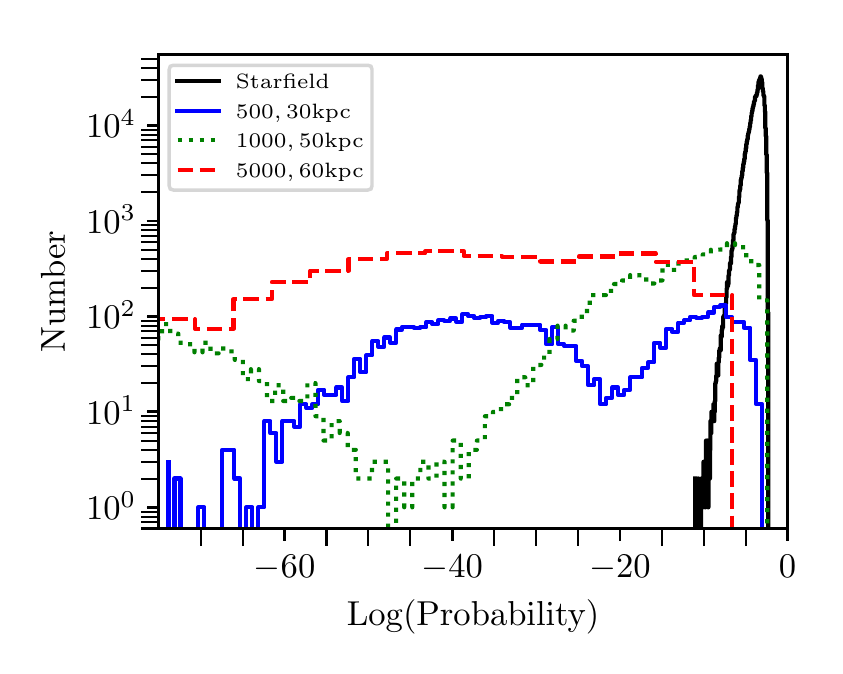}
\vspace{-3mm}
\caption{Log of the probability that the model stars (colored lines) are drawn from the parent distribution that describes the probability of the annuli {\bf A}$_x$ for a random field (black). The probabilities are calculated assuming that the counts within annuli are randomly distributed according to a Negative Binomial distribution. }
\label{fig:probs}
\vspace{1pt}
\end{figure}
%%%%%%%%%%%%%%%%%%%%%%%%%%%%%%%%%%%%%%%%%%

\subsection{Negative Binomial Model}
\label{sec:negbin}
% Lachlan's Idea for outlier detection

Our method relies on modeling the distribution of neighbor counts around each star in annuli (as described in \ref{subsec:gaia_count_catalog}) and looking for outliers, or low probability points, in this distribution. 

For each star, we model the vector {\bf A}, which is the neighbor counts in angular annuli on the sky over all scales up to 128\arcsec. We model each annulus $A_x$ in a small region of sky as an independent probability distribution, with no correlations between annuli. Then we have  that 
\begin{equation}
P(A | \phi) = \Pi_x P(A_x |\phi_x), 
\end{equation}
where $\phi_x$ are the model parameters. If all of the counts were random, we could model each probability distribution as a Poisson distribution. In practice, however, the dispersion may well be larger than the mean due to density variations over the fields of view that we consider. In order to account for this possibility, we specify $P(A_x)$ with the Negative Binomial distribution, which is a generalization of a Poisson distribution, and can be thought as a Poisson distribution whose mean is sampled from a gamma distribution. 

The Negative Binomial is a discrete probability distribution describing the number of `unsuccessful' trials $k$ that occur in some series of repeated, independent trials, each with success probability $p$, before $r$ successful trials occur. This distribution can be written as 
\begin{equation}
    \label{eq:neg_binom_def}
    p(k | r, p) = \binom{k+r-1}{r-1} p^r (1-p)^{k} \, . 
\end{equation}
This distribution approaches the Poisson distribution if one sends $ p\to 1$ and $r\to \infty$ while keeping the mean of the distribution [$\lambda = (1-p)r/p$] constant. Unlike in the Poisson distribution, the variance [$\sigma^2 = (1-p)r/p^2$] and mean are two independent quantities.

We determine the parameters of the Negative Binomial $\phi={r_x, p_x}$ by modeling the $A_x$ of stars in neighboring regions set to be a few times the expected scale of the hyper-compact clusters themselves.  We divide the sky into healpix pixels with {\tt NSIDE}$=256$ ($\sim 2$~\arcmin) pixels, and fit the Negative Binomial coefficients to stars in the surrounding nearest eight healpix pixels. We then use that parameter vector $\hat{\phi}$, comprising each fitted {$r_x,p_x$} for each annulus,  to estimate the likelihood of each star $P(A|\hat{\phi})$. We find $p$ to range from 0.7-0.9 at large radii, with significant improvements to $\chi^2$ over a Poisson fit. In the smallest annuli $<8$~\arcsec, $p$ approaches $1$, and we therefore adopt the Poisson distribution fits at these smallest annuli.

The final probability is calculated as the product of the probabilities from the fit to each annulus (Equation 2 above).  Stars that live in anomalously dense regions of sky on $\sim 15$ arcsec scales will have very low probability in this model, which assumes that stars are distributed randomly. In the next section we present mock tests that we use to pick an outlier threshold.

\subsection{Selecting outliers}
\label{sec:outlier}

To hone our outlier selection and test our completeness, we use the ArtPop models described in \S \ref{sec:artpop} to inject artificial clusters into our data, and calculate their Negative Binomial probabilities, as a function of number of stars in the cluster, distance to the cluster, and the background stellar density in each field. 

We have generated 100 realizations for each of three cluster masses ($N_{\rm star} = 500, 1000, 5000$~\msun) and two metallicities ([Fe/H]$=-1.5, -0.5$). These 100 models are each inserted into 100 random locations that span a range of background stellar density, for a total of $10^4$ models for each $N_{\rm star}$. For book-keeping purposes we track only the star closest to the center, creating annular counts that include both artificial stars in the cluster and real stars in the random location we have chosen. We then calculate the probability of each star in the Negative Binomial model. While these 100 models capture the variation in stars on the main sequence well, they will not capture the full stochasticity on the giant branch. Therefore we ran a second set of 1000 realizations for the $N=500, 1000$ low-metallicity models to verify that our completeness and thresholds do not change.

The distributions of probabilities from the Negative Binomial fits are shown in Figure \ref{fig:probs} for the lower-metallicity [Fe/H]$=-1.5$ model. For display purposes, we select models at a distance and number of stars that is roughly at our 80\% completeness limit, as shown in Figure \ref{fig:complete}. We then select a probability threshold for candidate clusters that retains a high fraction of the model cluster stars without swamping us with ``normal'' background stars. There are very few stars with probabilities log $P < -10$, but we can still maintain high completeness in the mock clusters (\S \ref{sec:completeness}). Generally, with this cut we select a fraction $\lesssim 0.01 \%$ of the stars, for a total of 22,200 candidates.

In principle, we also have access to proper motion data for the Gaia stars. We would expect the hyper-compact clusters to have velocities and positions consistent with halo objects, and we would expect the stars to have proper motions consistent with each other, as the internal motions of the cluster stars should be smaller than their bulk motions. However, in practice we are finding typically only 3-4 stars that may be associated with each other (\S \ref{sec:cmdfilter}), and some of these are too faint for reliable proper motions. In practice we therefore do not use the proper motion measurements in our search.

\subsubsection{Possible impact of crowding}
\label{sec:crowding}

Here we investigate what fraction of stars in the core of our clusters may be lost in the DECaLS catalog due to crowding (Fig.\ \ref{fig:modelclust}). While the clusters we are seeking will be a few arcsec across on the sky, as Figure \ref{fig:modelclust} shows, we could expect some blending to affect the number of stars in the DECaLS catalogs. While this blending will not impact our completeness corrections, that are based entirely on Gaia, they could impact our ability to detect the larger population of cluster stars that we expect at DECaLS depths.

We expect there to be 0.07 stars arcsec$^{-2}$ within the half-light radius of a 500-star cluster at our detection limit of $\sim 30$ kpc, 0.17 stars arcsec$^{-2}$ in a 1000-star cluster at our limit of $\sim 40$ kpc, and 0.9 stars arcsec$^{-2}$ in a 5000-star cluster at 50 kpc, beyond which we do not expect many clusters. We investigate how the DECaLS catalog behaves using the star clusters Reticulum and Whiting 1. Denser and more luminous clusters are modeled differently within the DECaLS catalog, which uses Gaia data to identify stars and does photometry only on these stars\footnote{ https:\///www.legacysurvey.org\//dr9\//external\//\#globular-clusters-planetary-nebulae}.

To get a handle on crowding using these clusters, we make two assumptions. First, we assume that all the objects at the centers of these two clusters are actually stars (probably nearly true). Second, we consider each object of any type to represent a single star. This is not always true. For instance, a single exponential object can be a blend of two or three stars. However, it will at least give us an idea of the magnitude of crowding. Whiting 1 has 0.09 stars arcsec$^{-2}$, with 70\% of those identified as point sources. Reticulum is denser, with 1.7 stars arcsec$^{-2}$ and 50\% of those identified as stars. Thus, at the crowding levels that we expect for the majority of our cases, it is safe to assume that $\sim 30\%$ of stars may be lost to crowding. Note that things will be worse in the center of a 5000-star cluster, but there would be many more stars outside the core that we would find. This is of course approximate, but it demonstrates that at the level of crowding that we expect, it is still safe to use the DECaLS photometry. We also note that we do not expect any saturated stars.

\subsection{Completeness}
\label{sec:completeness}

%%%%%%%%%%%%%%%%%%%%%%%%%%%%%%%%%%%%%%%%%%
\begin{figure} %[ht!]% 
\hspace{-5mm}
\includegraphics[width=8cm]{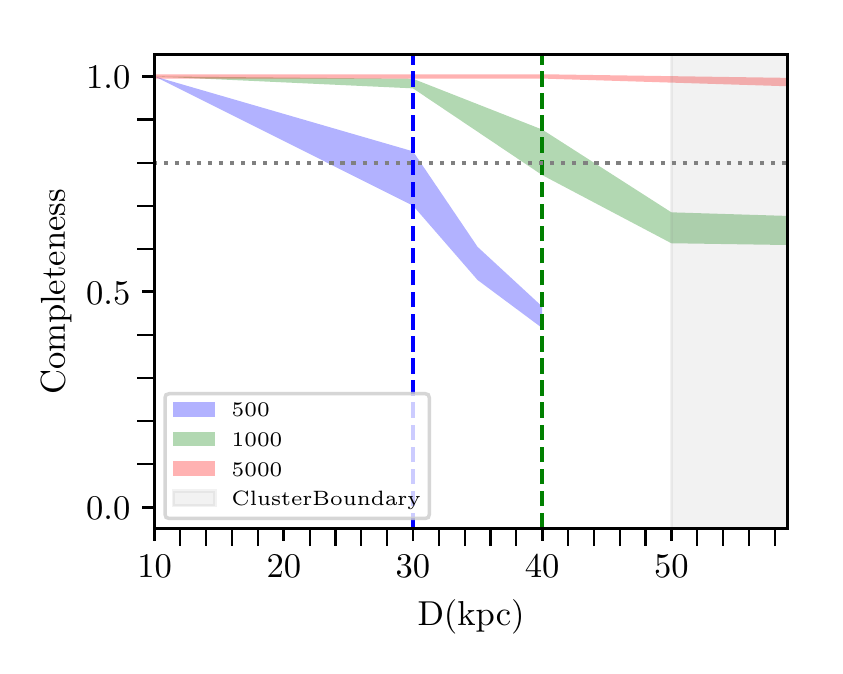}
\vspace{-2mm}
\caption{The fraction of clusters that we recover as outliers from our Negative Binomial fit for three cluster masses (500 stars in blue, 1000 stars green, 5000 stars red) as a function of distance. We define our survey volume by the distance where we reach 80\% completeness (dashed vertical blue for 500 stars, dashed vertical green for 1000 stars). Based on the {\tt SatGen} models, we expect most hyper-compact clusters would lie within 50 kpc of the Galactic Center, excluding the grey shaded region.}
\label{fig:complete}
\vspace{1pt}
\end{figure}
%%%%%%%%%%%%%%%%%%%%%%%%%%%%%%%%%%%%%%%%%%

We use the ArtPop simulations to quantify our completeness. For each cluster mass, and each distance, we first verify that there is no difference in completeness as a function of background stellar density in each field. This independence is not surprising, given that the search for very compact objects is limited by Poisson noise in the number of stars at the cluster scale rather than the Poisson noise in the background. We then measure the fraction of clusters that are identified by our outlier threshold based on the Negative Binomial fits, as a function of distance in Figure \ref{fig:complete}. Our probed volume is defined within the distance where we fall to 80\% completeness (30, 40 kpc for 500, 1000-star clusters, and beyond 50 kpc for 5000-star clusters). We will use these numbers to quantify the limits on the number of such clusters that may still be lurking in the Milky Way in \S \ref{sec:limits}.

\subsection{Eliminating candidates with DECaLS matching}
\label{sec:decalsfilter}

As shown in Figure \ref{fig:nstar}, we expect to detect many tens of stars from a cluster at our detection limit in DECaLS, where we might only expect four to six stars in Gaia. Therefore, a straightforward way to winnow down candidates identified with Gaia is simply to ask how many stars are detected at that position in DECaLS. From the DECaLS photometry, we count the number of stars and extract color-magnitude distributions, both of which can be used to determine whether or not we have found hints of a real cluster. 

By cross-matching with DECaLS, we are able to apply the following additional cuts. We only consider candidates that have at least $N_{\rm star, DECaLS}=9$ stars in the DECaLS catalog within 15 arcsec of the star in question. The aperture is chosen to be roughly twice the effective radius of a cluster around our completeness limit. We also apply a color criterion that the stars must fall in the range $0.2 < g-r < 1.5$, inspired by the color range of the color-magnitude diagram of an old and metal-poor stellar population (see Figure \ref{fig:modelclust}). Selecting nine stars as a limit is conservatively lower than the number we expect within $<15$ arcsec ($25 \pm 5$ at our detection limit for a 500-star cluster, Fig.\ \ref{fig:nstar}). Thus we allow for some loss of stars from confusion. However, with nine stars we still have sufficient numbers to crudely fit a color-magnitude relation (\S \ref{sec:cmdfilter}). 

As an additional sanity check on the impact of crowding, we also include in our ``star'' count objects that DECaLS has classified as ``compact'' exponential sources (REX). We only consider REX sources with sizes smaller than the PSF, but in this way we can crudely evaluate whether we are missing stars that have been classified as extended due to blending. In practice, we find that adding these makes little difference to our final list of targets, since most REX-dominated candidates tend to be distant galaxy clusters that are eliminated by our color cut. 

We find 176 stars from 86 candidate clusters with at least nine DECaLS stars falling in a reasonable color range, once we remove some contamination from inadequately masked nearby galaxies.  

We expect $\sim 25 \pm 5$ stars in a 15\arcsec\ aperture for the 500-star clusters at our detection limit (Fig. \ref{fig:nstar}). There is only one candidate cluster in our outlier sample with $N_{\rm star, DECaLS} > 12$, and 6 with $N_{\rm star, DECaLS} > 11$. This number is already $\sim 2 \sigma$ lower than the low end of the expected range at our distance detection limit, assuming Poisson errors. However, in order to determine whether we have identified any candidates, it is useful to also examine the colors and magnitudes of the candidates. With color and magnitude we can  isolate associated stars from foreground/background objects.

\subsection{Eliminating candidates with a color-magnitude fit}
\label{sec:cmdfilter}

%%%%%%%%%%%%%%%%%%%%%%%%%%%%%%%%%%%%%%%%%%
\begin{figure*} %[ht!]% 
\hspace{+1mm}
\begin{minipage}[b]{4cm}
\includegraphics[width=7cm]{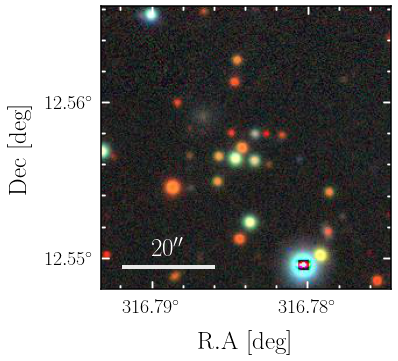}
\end{minipage}
\hspace{+30mm}
\begin{minipage}[b]{8cm}
\includegraphics[width=8.3cm]{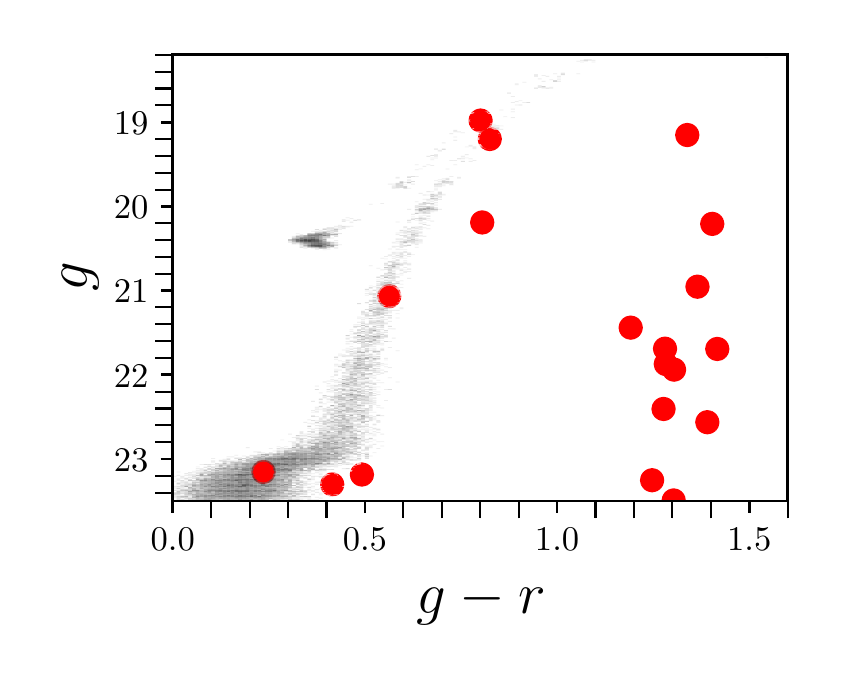}
\end{minipage}
\vspace{+2mm}
\caption{{\it Left}: A cutout from DECaLS around the position of the cluster candidate corresponding to the CMD at right.
{\it Right}: An example of the isochrone fit to the candidate cluster shown at left. The red points are drawn from a 30\arcsec\ radius around the candidate, while the grey-scale represents our best-fit isochrone. Even with this wider aperture, we still only find a maximum of six stars that may be associated with the same isochrone.}
\label{fig:CMDfit}
\vspace{1pt}
\end{figure*}
%%%%%%%%%%%%%%%%%%%%%%%%%%%%%%%%%%%%%%%%%%

For all cluster candidates, we next determine whether the DECaLS $gr$ color-magnitude distribution is consistent with an old coeval stellar population. We limit our attention to those candidates that have at least $N_{\rm star, DECaLS}=9$ stars with $0.2 < g-r < 1.5$ within 15\arcsec.

To fit the color-magnitude diagrams (CMDs), we include two components, a background model built from the data, and a simple stellar population model at a range of distances. 
The model of the background is constructed from all of the stars in 10 degree patches. The possible additional cluster component is made from the MESA Isochrones and Stellar Tracks isochrones \citep[MIST;][]{dotter2016,choietal2016} built from the Modules for Experiments in Stellar Astrophysics \citep[MESA][]{paxtonetal2011,paxtonetal2013,paxtonetal2015,paxtonetal2018} models sampled with a Kroupa IMF. For simplicity, and because we do not expect young stars in our hyper-compact star clusters, we fix the age to 10 Gyr and the metallicity to [Fe/H]$=-1.5$. There are two free parameters in our fit, the relative fraction of background and cluster stars, and the distance for the cluster component, allowed to range from 10 to 90 kpc.

We build a two-dimensional histogram of the theoretical color-magnitude diagram thus constructed, including a scatter in the magnitude and color derived from the DECaLS data. We then select the distance modulus and background weight that maximizes the likelihood (log $\mathcal{L} = \Sigma_i \rm{log} \rm{CMD}_i$), where CMD is the normalized probability density at the position of each of $i$ stars in our candidate cluster. An example fit is shown in Figure \ref{fig:CMDfit}.

We test this method using our mock clusters generated by ArtPop, described above in \S \ref{sec:artpop}. Specifically, we embed the model clusters in random DECaLS fields, add photometric noise, and then run our fit on this collection of stars. Once main sequence turnoff stars are too faint to be detected by DECaLS, we can no longer estimate a reliable distance (roughly $>40$~kpc), but we can achieve reasonable fits when the number of stars in the cluster is $N_{\rm star, DECaLS} > 9$. 

Our most likely CMD fits yield a refined $N_{\rm star, DECaLS}$ for each candidate cluster. Although we have identified overdensities in space, the colors of the stars suggests that the majority are foreground/background stars. According to our fits, the largest number of stars that may be associated with a common isochrone is $N_{\rm star, DECaLS} = 4-5$ stars (two candidates), or $N_{\rm star, DECaLS} = 3$ stars, (14), while all the other potential cluster stars are drawn from the background. Since the low end of our predicted number of stars is $25 \pm 5$ at our detection limit, we conclude that with high significance, we have not detected a candidate hyper-compact star cluster as defined in \S \ref{sec:prediction}.

As an example of an apparent overdensity, we highlight one candidate cluster from the list of 86 (Figure \ref{fig:CMDfit}). Within 15\arcsec, this candidate has three stars associated with the same isochrone, which grows to 5-6 when we open the aperture to 30\arcsec. This candidate comprises a clear overdensity spatially. As a sanity check, we fit Poisson distributions to randomly selected 15\arcsec\ apertures within 3\arcmin\ of the candidate, and find that it is a $3 \, \sigma$ outlier in density. However, as judged from the isochrone fit, only a few of those stars may be associated with each other. They do have consistent proper motions, but even if they are physically associated, there are not enough of them to represent the clusters that we are looking for.

\section{Limits on Wandering Black Holes in the Milky Way}
\label{sec:limits}

We searched $\sim 8000$ deg$^2$ of sky for hyper-compact star clusters of 500-5000 stars within 50 kpc of the Galactic Center, that would be the signature of wandering black holes accreted with infalling satellites. We did not find any plausible candidates, so in this section we translate our non-detection into upper limits on this population.

To calculate the upper limits, we assume that we are sensitive within the volume defined by our 80\% completeness, as indicated in Figure \ref{fig:complete}. We further assume that the black holes lie within 50 kpc, as calculated using the \msigma\ relation to populate halos with black holes (see Figure \ref{fig:radbh}). 

The resulting limits are shown in Figure \ref{fig:nbh-withlim} and Table \ref{table:limits}. The predicted number of black holes (or corresponding star clusters on the top axis) under the assumption of 100\% occupation fraction are shown by the dotted line, while taking the nuclear cluster occupation fraction as the black hole occupation fraction is shown in solid. In what follows we discuss what we can and cannot conclude from these limits.

\subsection{Limits on the occupation fraction}

In this paper we have searched for and place limits on the number density of star clusters with $< 5000$ stars. We are interested in the corresponding number of black holes. We translate between the cluster mass limit and \mbh\ by simply assuming that the cluster mass is 20\% of \mbh.

The number of black holes that we expect is basically the product of the halo mass function, the occupation fraction, and a convolution with the halo to black hole scaling relation. If we assume that \sigmastar\ can be related to the maximum halo circular velocity as \sigmastar$ = v_{\rm max}/2.2$, then we can take the black hole scaling relation with $v_{\rm max}$ to have the form: log \mbh $= \alpha + \beta {\rm log} \, v_{\rm max}$. For intrinsic scatter in this relation of $\sigma_{\rm int}$, we then have \citep[see, e.g.,][]{marconietal2004,gallosesana2019}:

\begin{equation}
  \begin{split}
  P(M_{\rm BH} | v_{\rm max}) & =  \\
  & \frac{1}{\sqrt{2 \pi} \sigma_{\rm int}} \exp{ -\frac{1}{2} \left[ \frac{\log M_{\rm BH} - \alpha - \beta \log v_{\rm max}}{2 \sigma_{\rm int}} \right]^2  }
  \end{split}
\end{equation}

\noindent
We then write the number of black holes in terms of the halo maximum velocity function, the occupation fraction $\lambda_{\rm occ}(v_{\rm max})$, and $P(\log M_{\rm BH} | v_{\rm max})$:

\begin{equation}
    N(M_{\rm BH}) = 
    \int \frac{1}{M_{\rm BH}} N_h(v_{\rm max}) \lambda_{\rm occ}(v_{\rm max}) P(M_{\rm BH} | v_{\rm max}) \, d v_{\rm max}  
\end{equation}

\noindent
where the halo velocity function is given by $N_h(v_{\rm max})$ and the occupation fraction is $\lambda_{\rm occ}$ ($v_{\rm max}$). 

The halo mass function in our predictions is derived from the {\tt SatGen} simulation. The \mbh\ scaling relation is extrapolated from the observed \msigma\ relation, assuming that the maximum halo circular velocity can be converted directly to a stellar velocity dispersion with a constant scale factor. Both the extrapolation and the conversion are uncertain. As described in \S \ref{sec:satgen}, we do not consider the $M_*-$\mbh\ scaling relation here, given the additional uncertainty added by the unknown stellar-to-halo mass relation.

Currently, the occupation fraction is not constrained in the regime of interest to us, below galaxy stellar masses of $M_* \approx 10^9$~\msun, or \sigmastar$\approx 20$~km~s$^{-1}$. There are observational constraints on occupation fraction for more massive dwarf galaxies with stellar masses of $M_* \gtrsim 10^9$~\msun\ \cite[see summary in][]{greeneetal2020}, where the occupation fraction is consistent with at least 50\% of galaxies hosting a central black hole, based both on stellar dynamical results \citep[e.g.,][]{nguyenetal2018,nguyenetal2019} and X-ray studies \citep[e.g.,][]{milleretal2015,sheetal2017}. At stellar masses $M_* < 10^9$~\msun, there are precious few constraints on the occupation fraction, although some AGN have been found in galaxies of this mass \citep[e.g.,][]{baldassareetal2019,reinesetal2019}. 

Our measured limits are inconsistent with 100\% occupation for sub-halos of the Milky Way. They are well below predictions for the number of black holes we would expect if every satellite contained a black hole with mass as predicted by the \msigma\ scaling. As mentioned in \S \ref{sec:satgen}, a heavy-seeding model in which all halos down to $\sim 10^6$~\msun\ are seeded is ruled out even more conclusively.

\begin{figure} %[ht!]% 
\hspace{-5mm}
\includegraphics[width=9cm]{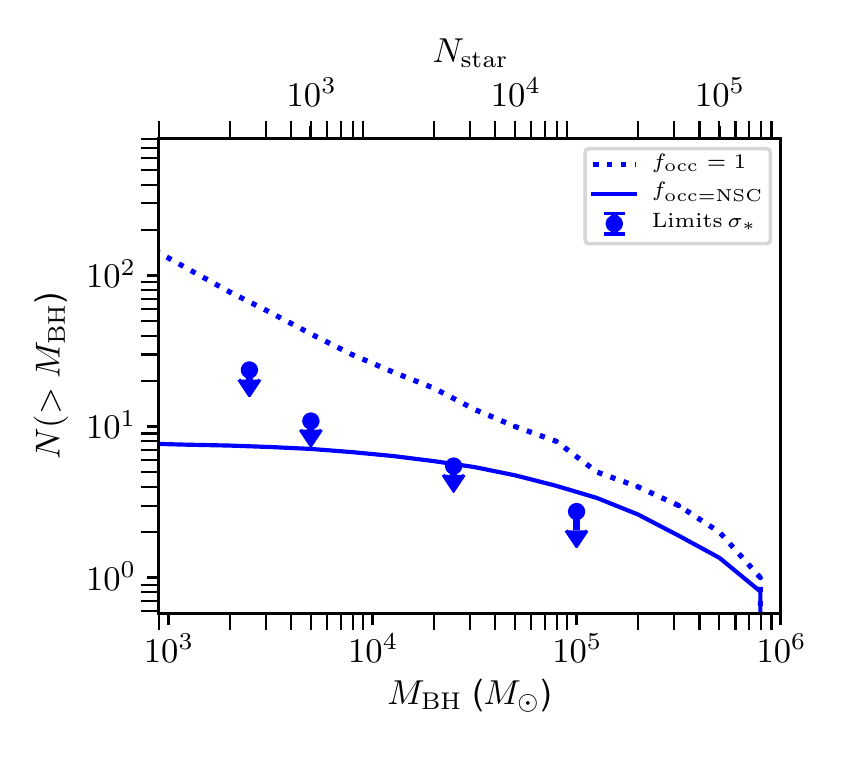}
\vspace{-2mm}
\caption{Upper limits on the number of hyper-compact star clusters in the Milky Way halo within 50 kpc. Theoretical expectations from {\tt SatGen}, assuming either that every satellite hosts a $10^3-10^5$~\msun\ black hole are shown in dotted, or in solid the perhaps more realistical case that the fraction of galaxies hosting black holes is the same as the fraction of galaxies that host nuclear star clusters. Black hole masses are estimated from the halo maximum velocity (blue). Our limits assume that all clusters would be found within 50 kpc of the Galactic Center.}
\label{fig:nbh-withlim}
\vspace{1pt}
\end{figure}

To take a concrete example of a non-unity occupation fraction, we adopt the nucleation fraction (the fraction of galaxies containing nuclear star clusters) as a proxy for the black hole occupation fraction, shown as a solid line in Fig.\ \ref{fig:nbh-withlim}). All black holes discovered dynamically in galaxies with $M_* \approx 10^9-10^{10}$~\msun\ are found in nuclear star clusters \citep[e.g.,][]{sethetal2008,nguyenetal2018,nguyenetal2019}, potentially suggesting a relationship between the two \citep[see details in ][]{neumayeretal2020}. Nucleation fractions are near unity at $\sim 10^{9.5}$~\msun, and then fall to $<10\%$ by $M_* \sim 10^6$~\msun\ \citep[e.g.,][]{sanchez-janssenetal2019,neumayeretal2020}. 

Our observed limits are consistent with an occupation fraction that falls with mass in a similar way to the nucleation fraction. In fact, the details of how we seed the satellites become relatively unimportant in this case, because the number of black holes is set by the small number of available dissolved satellites with initial $M_* > 10^7$~\msun, where most of the black holes would be found. If we improve the probed area by a factor of a few (e.g., with the Rubin Observatory; \citealt{ivezicetal2019}), we might be able to constrain the occupation fraction mass dependence further. In the meantime, we still expect that there must be a wandering black hole population in Milky-Way like galaxies, and we explore other ways to search for them in \S \ref{sec:othersearch}.

%%%%%%%%%%%%%%%%%%%%%%%%%%%%%%%%%%%%%%%%%%
\begin{deluxetable}{lll}
\tablecolumns{3}
\tablewidth{0pt}
\tablecaption{Number Density Limits \label{table:agnfrac}}
\tablehead{
\colhead{$N_{\rm star}$} & \colhead{$M_{\rm BH} (\msun)$} & \colhead{$N(<M_{\rm BH})$}}
\startdata
$5 \times 10^2$ & $10^3$ &  24  \\
$10^3$ & $5 \times 10^3$ &  11  \\
$5 \times 10^3$ & $10^4$ &  5  \\
$10^4$ & $5 \times 10^4$ &  3  \\
%%%%%%%%%%%%%%%%%%%%%%%%%%%%%%%
\enddata
\tablecomments{Limits on number of hyper-compact star clusters, and corresponding limits on the number of wandering massive black holes inferred in this work.}
\label{table:limits}
\end{deluxetable}

%%%%%%%%%%%%%%%%%%%%%%%%%%%%%%%%%%%%%%%%%%

%%%%%%%%%%%%%%%%%%%%%%%%%%%%%%%%%%%%%%%%%%
%%%%%%%%%%%%%%%%%%%%%%%%%%%%%%%%%%%%%%%%%%

\subsection{Important limitations to our assumptions}

Dynamical modeling that resolves the scale of the sphere of influence of the black hole is needed to be secure in the properties of the hyper-compact star clusters. There are at least three effects that we have ignored or oversimplified that could strongly impact our conclusions. First, we assume that once a satellite reaches a stellar mass of $10^6$~\msun\ in {\tt SatGen}, it will fully dissolve. This assumption should be tested in the context of a central intermediate-mass black hole. 

Second, we have ignored the possibility that hyper-compact star clusters may be composed mostly of compact remnants. Given that neutron stars and black holes will settle to the cluster center through mass-segregation, there is some chance that these clusters exist but have much higher mass-to-light ratios than assumed here. In fact, a mass fraction in stellar-mass remnants of $\sim 20\%$, which sounds high compared with some models \citep[e.g.,][]{zocchietal2019,baumgardtetal2019}, has recently been suggested to explain the low-density of Palomar 5 and potentially other fluffy globular clusters \citep{gielesetal2021}. It would be very interesting to know how the black hole pathways investigated by Gieles et al.\ play out with an intermediate-mass black hole at the cluster center. 

Third, we have assumed that all of the clusters have the same size of 1 pc, which is clearly an oversimplification and is also related to the detailed dynamical evolution of the cluster. If the clusters are significantly larger than the $\sim 1$~pc that we assume here, then their clustered signal would drop, as would our sensitivity. As an example, if the $N_{\rm star} = 1000$ cluster had a size of 5~pc, then our completeness would drop roughly by a factor of two compared to what is shown in Figure \ref{fig:complete}.

\subsection{Other ways to search for wandering black holes}
\label{sec:othersearch}

While we have been focused on low-mass hyper-compact clusters, one promising avenue for continued study is certainly to identify all of the Milky Way star clusters that may be stripped nuclei, using both stellar population and orbital information \citep[e.g.,][]{massarietal2019,pfefferetal2021}. Future observations with extremely large telescopes will provide far more stringent limits on their possible central black hole content.

Another angle is to think about the other channels that may generate hyper-compact clusters surrounding black holes at larger galactocentric radius, including early ejection from the gravitational slingshot \citep[e.g.,][]{volonterietal2003,olearyloeb2009,lenaetal2020}. It is therefore interesting to consider searches that will reach larger galactocentric radii in the Milky Way, as well as complementary techniques for searching for extragalactic analogs. 

Euclid and the Vera Rubin Observatories Legacy Survey of Space and Time \citep[LSST;][]{ivezicetal2019} will provide a more sensitive search of these potentially more distant hyper-compact star clusters. Here we consider what these clusters would look like in next-generation imaging surveys. From the ground, the LSST will produce single-epoch images with $5 \,\sigma$ depths of $r = 24.7$~mag. At this depth, we could expect to uncover 500-star hyper-compact clusters out to $\sim 70$~kpc, where we still expect $\sim 25 \pm 5$ stars to be detected within a 3\arcsec\ radius, although crowding will likely be significant. A 1000-star cluster will have $r = 17-18$ mag within $\sim 2$\arcsec\ at 70-90 kpc, making them easily detectable in principle. However, these clusters will be very crowded at ground-based spatial resolution, and so techniques using colors may be required for these more massive clusters beyond $\sim 50$~kpc. Such clusters may be selected as outliers in the color-magnitude diagram, as they will be far more luminous than expected for their color \citep[e.g.,][]{lenaetal2020}. One might imagine combining such a ground-based search with higher-resolution data from a mission like Euclid \citep{raccaetal2016} to see if the putative clusters are resolved into a couple of stars at higher resolution.

Finally, an additional prospect is opened by the Roman Space Telescope \citep{spergeletal2015} to search for the $\sim 5000$ star hyper-compact clusters in halos of external galaxies. While these clusters would be point sources, they would be anomalously bright for their color. Since such a cluster would have an integrated magnitude of $z \approx 22-23$ mag at 10 Mpc, one could imagine performing a search leveraging the proposed Roman Infrared Nearby Galaxy Survey (RINGS) \citep{wings15} to search for hyper-compact star clusters at the very low-luminosity end of the star cluster luminosity function, using the high spatial resolution to remove background galaxies and the colors to distinguish from halo stars, as extragalactic globular clusters are currently found. However, ultimately identifying these as real cluster candidates will be very challenging, as will certainly require spectroscopy.

Another tool, and the most robust one, is to search for the dynamical signatures of black holes in the motions of stars. At somewhat higher mass, stripped remnant clusters surrounding massive black holes have been found from dynamical signatures in ultra-compact dwarf galaxies \citep{sethetal2014,ahnetal2017,ahnetal2018}. Closer to home in Andromeda, lower-mass stellar clusters, likely to be ultra-compact dwarfs, also show some signs of nuclear black holes \citep{gebhardtetal2005} although these remain very challenging measurements. In the era of extremely large telescopes, we can hope to detect $10^4$~\msun\ black holes dynamically out to $\sim 5-10$ Mpc, which will reach a much larger number of massive globular cluster and UCDs \citep{greeneetal2019}. 

There are promising search avenues via accretion signatures as well. One is through tidal capture of individual stars in the cluster \citep[e.g.,][]{macleodetal2016imbh}, which may be what we are observing in the hyper-luminous X-ray source HLX1 \citep[e.g.,][]{farrelletal2009,lasotaetal2011,soriaetal2017}. Tidal disruption events should probe wandering black hole populations, and there are some potential tidal events in off-nuclear stellar systems \citep[e.g.,][]{linetal2018}.

It is possible that wandering black holes in more massive halos (e.g., the Virgo cluster) may encounter sufficient gas to be detected via accretion of inter-cluster medium \citep[e.g.,][]{guoetal2020}. This detection channel is highly unlikely in Milky Way-mass systems. It may, however, be possible to detect very low Eddington-ratio accretion onto black holes in the centers of normal extragalactic globular clusters in the radio, which is relatively boosted at low accretion rates \citep[e.g.,][]{maccaroneetal2005}. On the other hand, searches with deep radio surveys have not been successful to date, either in the Milky Way \citep{tremouetal2018} or external galaxies \citep{wrobeletal2016}. Next-generation radio facilities will be very powerful to further the reach of these searches \citep{wrobeletal2019}.

Finally, wandering black holes may eventually be detected as extreme mass-ratio inspiral events through the merger of a stellar mass black hole with the central intermediate-mass black hole \citep[e.g.,][]{gairetal2010,amaro-seoaneetal2015,eracleousetal2019}. Using the tools developed here, it would be useful to calculate the number densities of extragalactic wandering black holes to estimate a detection rate for space-based gravitational wave experiments.

\section{Summary}
\label{sec:summary}

We have presented a search for hyper-compact star clusters that would be the hosts of wandering black holes. If such black holes are deposited by infalling satellites, we expect them to be $\sim 1$~pc in size, have 500-5000 stars, and fall within $\sim 50$~kpc of the Galactic Center. We use Gaia+DECaLS data over $\sim 8000$ deg$^2$. Using a Negative Binomial model to describe the distributions of stellar counts around each target star in Gaia EDR3, we identify large outliers in count space. Real stellar clusters with normal mass functions would have three-five times as many stars in DECaLS, allowing us to efficiently eliminate candidates through a cross-match with DECaLS. We do not find any hyper-compact star clusters within $\sim 30-50$~kpc (for clusters of 500-5000~\msun). We translate these limits into upper limits on the number of intermediate-mass black holes wandering in the inner Milky Way halo. 

We also use modern semi-analytic models to bracket the number of $10^3-10^5$~\msun\ black holes that might wander in the Milky Way halo using the {\tt SatGen} code \citep{jiangetal2019}. {\tt SatGen} calculates the number of expected dissolved satellites as a function of galactocentric radius and halo/stellar mass at infall. We then extrapolate black hole-galaxy scaling relations to predict the range of black hole masses that might inhabit these halos. Based on our calculations, we expect as many as 100 black holes with \mbh$> 10^3$~\msun\ based on scaling from the \msigma\ relation, should every satellite carry an intermediate-mass black hole. In the context of this model, we can rule out that all satellite galaxies host central black holes from our non-detection. Our measurements are consistent with models in which most black hole hosts have stellar masses $>10^7$~\msun.

In the near future, wide-area imaging surveys from the ground and space will open up new discovery space for hyper-compact star clusters in the Milky Way and nearby galaxies. Extremely large telescopes will enable dynamical detection of $10^3$~\msun\ black holes in the Milky Way and $10^4$~\msun\ black holes within the Local Volume. The Rubin Observatory LSST will increase the rate of detection of tidal disruption events, some of which may be in off-nuclear systems. Eventually gravitational wave detectors in space will be sensitive to the mergers of stellar-mass and intermediate-mass black holes in these clusters \citep[e.g.,][]{gairetal2010,gallosesana2019}. Thus, if there are black holes wandering in galaxy halos, we will begin to uncover them relatively soon.

\acknowledgments
JEG acknowledges support from NSF grants AST-1815417 and AST-1907723. YST is grateful to be supported by the NASA Hubble Fellowship grant HST-HF2-51425.001 awarded by the Space Telescope Science Institute. S.D. is supported by NASA through Hubble Fellowship grant HST-HF2-51454.001- A awarded by the Space Telescope Science Institute, which is operated by the Association of Universities for Research in Astronomy, Incorporated, under NASA contract NAS5-26555. J.P.G. is supported by an NSF Astronomy and Astrophysics Postdoctoral Fellowship under award AST-1801921. 
J.S. acknowledges support from NSF grant AST-1812856 and the Packard Foundation. This paper made use of the Whole Sky Database (wsdb) created by Sergey Koposov and maintained at the Institute of Astronomy, Cambridge by Sergey Koposov, Vasily Belokurov and Wyn Evans with financial support from the Science \& Technology Facilities Council (STFC) and the European Research Council (ERC).

The Legacy Surveys consist of three individual and complementary projects: the Dark Energy Camera Legacy Survey (DECaLS; Proposal ID \#2014B-0404; PIs: David Schlegel and Arjun Dey), the Beijing-Arizona Sky Survey (BASS; NOAO Prop. ID \#2015A-0801; PIs: Zhou Xu and Xiaohui Fan), and the Mayall z-band Legacy Survey (MzLS; Prop. ID \#2016A-0453; PI: Arjun Dey). DECaLS, BASS and MzLS together include data obtained, respectively, at the Blanco telescope, Cerro Tololo Inter-American Observatory, NSF's NOIRLab; the Bok telescope, Steward Observatory, University of Arizona; and the Mayall telescope, Kitt Peak National Observatory, NOIRLab. The Legacy Surveys project is honored to be permitted to conduct astronomical research on Iolkam Du'ag (Kitt Peak), a mountain with particular significance to the Tohono O'odham Nation.

NOIRLab is operated by the Association of Universities for Research in Astronomy (AURA) under a cooperative agreement with the National Science Foundation.

This project used data obtained with the Dark Energy Camera (DECam), which was constructed by the Dark Energy Survey (DES) collaboration. Funding for the DES Projects has been provided by the U.S. Department of Energy, the U.S. National Science Foundation, the Ministry of Science and Education of Spain, the Science and Technology Facilities Council of the United Kingdom, the Higher Education Funding Council for England, the National Center for Supercomputing Applications at the University of Illinois at Urbana-Champaign, the Kavli Institute of Cosmological Physics at the University of Chicago, Center for Cosmology and Astro-Particle Physics at the Ohio State University, the Mitchell Institute for Fundamental Physics and Astronomy at Texas A\&M University, Financiadora de Estudos e Projetos, Fundacao Carlos Chagas Filho de Amparo, Financiadora de Estudos e Projetos, Fundacao Carlos Chagas Filho de Amparo a Pesquisa do Estado do Rio de Janeiro, Conselho Nacional de Desenvolvimento Cientifico e Tecnologico and the Ministerio da Ciencia, Tecnologia e Inovacao, the Deutsche Forschungsgemeinschaft and the Collaborating Institutions in the Dark Energy Survey. The Collaborating Institutions are Argonne National Laboratory, the University of California at Santa Cruz, the University of Cambridge, Centro de Investigaciones Energeticas, Medioambientales y Tecnologicas-Madrid, the University of Chicago, University College London, the DES-Brazil Consortium, the University of Edinburgh, the Eidgenossische Technische Hochschule (ETH) Zurich, Fermi National Accelerator Laboratory, the University of Illinois at Urbana-Champaign, the Institut de Ciencies de l'Espai (IEEC/CSIC), the Institut de Fisica d'Altes Energies, Lawrence Berkeley National Laboratory, the Ludwig Maximilians Universitat Munchen and the associated Excellence Cluster Universe, the University of Michigan, NSF's NOIRLab, the University of Nottingham, the Ohio State University, the University of Pennsylvania, the University of Portsmouth, SLAC National Accelerator Laboratory, Stanford University, the University of Sussex, and Texas A\&M University.

The Legacy Surveys imaging of the DESI footprint is supported by the Director, Office of Science, Office of High Energy Physics of the U.S. Department of Energy under Contract No. DE-AC02-05CH1123, by the National Energy Research Scientific Computing Center, a DOE Office of Science User Facility under the same contract; and by the U.S. National Science Foundation, Division of Astronomical Sciences under Contract No. AST-0950945 to NOAO.

\bibliography{blackhole.bib}

\end{document}